\ifdefined\pdfoutput\ifx\XeTeXversion\undefined\pdfoutput=1\fi\fi
\documentclass[11pt,a4paper]{article}

\usepackage[utf8]{inputenc}
\usepackage[T1]{fontenc}
\usepackage{lmodern}
\usepackage[margin=2.3cm]{geometry}
\usepackage{amsmath,amssymb,amsfonts,amsthm,mathtools}
\usepackage{booktabs,array,tabularx}
\usepackage{graphicx}
\graphicspath{{figures/}{../figures/}{./}}
\usepackage{subcaption}
\usepackage{microtype}
\usepackage{xcolor}
\usepackage{hyperref}
\usepackage{algorithm}
\usepackage{algpseudocode}
\usepackage{cite}

\hypersetup{
    colorlinks=true,
    linkcolor=blue!70!black,
    citecolor=green!50!black,
    urlcolor=blue!80!black
}

\newtheorem{theorem}{Theorem}
\newtheorem{lemma}{Lemma}
\newtheorem{proposition}{Proposition}
\newtheorem{corollary}{Corollary}

\newtheorem{remark}{Remark}

\DeclareMathOperator{\SDiff}{SDiff}
\DeclareMathOperator{\Diff}{Diff}

\DeclareMathOperator{\Vol}{Vol}

\DeclareMathOperator{\diag}{diag}
\DeclareMathOperator{\clip}{clip}

\title{\textbf{Exact Volume-Preserving Diffeomorphic Shear Splitting:\\Continuous-Discrete Geometric Bounds, Reversible Memory Backpropagation, and Expressivity Frontiers in Volumetric Mapping}}

\author{
    \textbf{{\'A}lvaro Garc{\'i}a-Casas}\footnote{Corresponding author. E-mail: \texttt{alvaro.g.casas@alumnos.uc3m.es}}\\[0.5em]
    \small Department of Bioengineering, Universidad Carlos III de Madrid\\
    \small Avenida de la Universidad 30, 28911 Legan{\'e}s (Madrid), Spain
}

\date{\today}

\begin{document}

\maketitle

\begin{abstract}
\noindent \textbf{Purpose:} Incompressible diffeomorphic image registration requires estimating continuous spatial mappings that preserve local tissue volume ($\det(D\Phi) \equiv 1$) without topological folding ($\det(D\Phi) > 0$). Classical methods based on divergence-free stationary velocity fields (SVF) or Large Deformation Diffeomorphic Metric Mapping (LDDMM) approximate continuous incompressibility through numerical ODE integration, scaling-and-squaring, or soft divergence penalties. These approaches exhibit numerical volume drift ($\sim 10^{-4} - 10^{-1}$), require iterative grid interpolations for inversion, and incur linear backpropagation activation memory ($\mathcal{O}(L \cdot N_{\text{vox}})$).

\noindent \textbf{Methods:} We formulate a structure-preserving geometric framework for 3D volumetric registration based on cascaded planar coordinate shears. By decomposing the divergence-free Lie algebra $\mathfrak{sdiff}(\Omega)$ into six directional coordinate-shear generators whose convective non-linearities vanish identically, each elementary stage integrates analytically to a closed-form triangular shear with $\det(D\mathcal{S}_k) \equiv 1$ and exact machine-precision inverse $\mathcal{S}_k^{-1} = \mathcal{S}_k(-\Delta t)$ requiring zero fixed-point iterations and zero spatial interpolations. Through a two-step finite-to-infinite-dimensional spectral Galerkin argument and the Baker--Campbell--Hausdorff expansion, multi-stage compositions generate the connected volume-preserving group $\operatorname{SDiff}_0(\Omega)$. We establish: (i) an unconditional global diffeomorphism proof closing the Hadamard--Caccioppoli injectivity gap; (ii) a scaled-support Region of Interest (ROI) Dirichlet confinement lemma guaranteeing zero boundary escape; (iii) an anisotropic Riemannian metric similarity invariance; (iv) a Sobolev $H^2$ spectral Lipschitz budget preventing chaotic phase-space mixing (the 3D continuous Arnold cat-map analogue); and (v) a custom reversible backpropagation engine eliminating $\mathcal{O}(L)$ intermediate trajectory storage ($0.00\text{ MB}$ intermediate graph activations).

\noindent \textbf{Results:} We prove and numerically verify second-order $\mathcal{O}(h^2)$ sub-voxel convergence on canonical Kuhn 6-tetrahedral partitions across $16^3$ to $64^3$ lattices (OLS log-log slope $2.217$, $24.0\times$ error reduction, $0.0000\%$ inverted tetrahedra) and second-order $\mathcal{O}(\Delta t^2)$ trajectory accuracy for a 12-stage palindromic Strang splitting (OLS slope $2.001$, mean-error reduction factor $468.1\times$ and per-instance ratio $504.4 \pm 80.4\times$ over Lie--Trotter at $L=16$). Across $32^3$ to $128^3$ volumes, forward-inverse execution is $2.1\times - 4.0\times$ faster than SVF scaling-and-squaring with a $26.9\times$ reduction in peak backpropagation memory ($768.2\text{ MB}$ vs.\ $20,640.1\text{ MB}$ at $128^3, L=10$), while running $30.4\% - 64.6\%$ faster than PyTorch gradient checkpointing (\texttt{torch.utils.checkpoint}, $\text{Rev}/\text{Ckpt} = 0.354\times - 0.696\times$). On $N=20$ clinical cardiac cine MRI subjects (ACDC), we derive an analytical volume-preserving Dice upper bound of $0.9600 \pm 0.0309$ ($\Delta V = 8.48\% \pm 6.75\%$) and show via direct shape-oracle optimization ($N=5$ subgroup) that triangular shears achieve Dice $0.7622 \pm 0.0432$ and non-parametric curl potentials achieve $0.9498 \pm 0.0185$. Finally, we formalize the \textit{Volume-Fraction \& Global-Support Optimization Landscape Barrier}: unsupervised global intensity MSE stalls on clinical cine ($0.5202 \pm 0.1414$ vs.\ $0.5133 \pm 0.1413$ unregistered) because the myocardium occupies only $\gamma_{\text{ROI}} = 3.12\%$ of the volume while global Fourier modes couple across the $96.88\%$ static background.

\noindent \textbf{Conclusion:} Exact symplectic shear splitting provides a mathematically rigorous, unconditionally volume-preserving, and reversible computational foundation for 3D volumetric mapping, delineating the exact boundary between continuous geometric expressivity and non-convex intensity optimization landscapes.

\vspace{0.5em}
\noindent \textbf{Keywords:} Volume-preserving diffeomorphisms, Symplectic shear splitting, Kuhn tetrahedral mesh convergence, Reversible backpropagation, Arnold cat map, Mathematical imaging.
\end{abstract}

\newpage

\section{Introduction}
\label{sec:intro}

Deformable image registration is a foundational inverse problem in mathematical imaging, computer vision, and computational anatomy \cite{beg2005computing,ashburner2007fast,sotiras2013deformable}. In continuum biomechanics, soft biological tissues---such as myocardial muscle, hepatic parenchyma, and cerebral white matter---behave as nearly incompressible media over physiological deformation timescales \cite{rohlfing2003volume,bistoquet2008myocardial,mansi2011ilogdemons}. A physically admissible spatial transformation $\Phi: \Omega \to \Omega$ on a bounded domain $\Omega = [-1, 1]^3 \subset \mathbb{R}^3$ must therefore satisfy two strict geometric invariants:
\begin{enumerate}
    \item \textbf{Global Diffeomorphic Bijectivity:} $\Phi$ must be a smooth bijection with a smooth inverse ($\Phi \in \Diff^2(\Omega)$), precluding topological singularities, tearing, or local coordinate folding ($\det(D\Phi(\mathbf{x})) > 0$ for all $\mathbf{x} \in \Omega$).
    \item \textbf{Pointwise Volume Preservation (Incompressibility):} The continuous Jacobian determinant must equal unity everywhere:
    \begin{equation}
        \det\!\big(D\Phi(\mathbf{x})\big) \equiv 1 \qquad \forall \, \mathbf{x} \in \Omega.
        \label{eq:det_one}
    \end{equation}
\end{enumerate}
The set of such transformations forms the infinite-dimensional Fr\'echet Lie group $\SDiff(\Omega)$ of volume-preserving diffeomorphisms \cite{arnold1966geometrie,ebin1970groups}.

\subsection{Limitations of Existing Registration Paradigms}
Existing computational frameworks approximate continuous incompressibility only indirectly, incurring fundamental trade-offs between geometric exactness, invertibility, and memory complexity:
\begin{itemize}
    \item \textbf{Soft Divergence and Jacobian Penalization:} In B-spline Free-Form Deformations (FFD) \cite{rohlfing2003volume}, nearly incompressible biventricular models \cite{bistoquet2008myocardial}, and unconstrained dense displacement models, incompressibility is relaxed into a soft penalty $\lambda_{\text{div}} \int_\Omega (\det D\Phi - 1)^2 d\mathbf{x}$. Because the optimizer trades intensity similarity against volume conservation, finite weights permit substantial local volume drift ($|\det D_h\Phi - 1|_\infty \sim 10^{-1} - 10^0$), while large weights stiffen the optimization landscape.
    \item \textbf{Scaling-and-Squaring of Stationary Velocity Fields (SVF):} In log-Euclidean frameworks such as iLogDemons \cite{mansi2011ilogdemons} and spectral SVF, a divergence-free velocity field $\mathbf{v}$ ($\nabla \cdot \mathbf{v} = 0$) is integrated via $K$ recursive scaling-and-squaring steps: $\Phi = \exp(\mathbf{v}) \approx (\text{Id} + 2^{-K}\mathbf{v})^{\circ 2^K}$. Although continuous integration of a solenoidal field preserves volume, repeated trilinear grid interpolations at each squaring stage introduce cumulative Jacobian drift ($\sim 10^{-4} - 10^{-2}$) and inversion round-trip error ($\|\Phi^{-1}(\Phi(\mathbf{x})) - \mathbf{x}\|_\infty \sim 10^{-4} - 10^{-3}$).
    \item \textbf{Eulerian ODE Integration (LDDMM):} In Large Deformation Diffeomorphic Metric Mapping \cite{beg2005computing,vialard2012diffeomorphic}, integrating time-dependent velocity fields via explicit Runge--Kutta (RK4) schemes requires storing all intermediate 3D coordinate grids across $L$ integration steps during reverse-mode automatic differentiation, incurring $\mathcal{O}(L \cdot N_{\text{vox}})$ memory overhead that exceeds $20\text{ GB}$ at $128^3$ resolution in double precision.
\end{itemize}

\subsection{Distinction from Euclidean Normalizing Flows and Reversible Networks}
Triangular coupling layers with unit Jacobian determinant were introduced in generative density estimation by Dinh et al.\ (NICE \cite{dinh2014nice}, RealNVP \cite{dinh2017density}), and reversible activation reconstruction was introduced for residual classification networks by Gomez et al.\ (RevNet \cite{gomez2017reversible}). However, directly applying unrestricted Euclidean coupling layers to 3D volumetric continuum mapping on bounded physical domains fails for four mathematical reasons that we resolve in this work:
\begin{enumerate}
    \item \textbf{Boundary Escape on Bounded Domains:} Standard NICE/RealNVP layers act on $\mathbb{R}^d$ without boundary conditions. On a bounded imaging volume $\Omega = [-1,1]^3$, an unwindowed shear $x_{i_k} \mapsto x_{i_k} + f_k(u,v)$ pushes anatomical voxels across $\partial\Omega$. Naive 3D Dirichlet windowing $w(x_1,x_2,x_3)\tilde{f}_k$ introduces dependence on the active coordinate $x_{i_k}$, destroying $\det(D\mathcal{S}_k) \equiv 1$ and closed-form invertibility. Our Lemma~\ref{lem:boundary_and_metric} resolves this via a scaled-support passive cosine envelope $w_\delta(u,v)$ coupled with a Courant--Friedrichs--Lewy (CFL) step-size bound.
    \item \textbf{Anisotropic Physical Metric Tensors:} Medical scans have anisotropic physical voxel dimensions $\mathbf{S} = \diag(s_1, s_2, s_3)$ (e.g., $1.4 \times 1.4 \times 5.0\text{ mm}$). We establish metric similarity invariance (Lemma~\ref{lem:boundary_and_metric}) and scale-compensated stream potentials on physical manifolds.
    \item \textbf{Continuous-to-Discrete Mesh Truncation Gap:} Generative normalizing flows evaluate densities at isolated point samples, whereas volumetric registration warps continuous grids interpolated over cubic lattices $\Omega_h$. We prove that while $\det(D\Phi_\theta) \equiv 1$ holds to machine precision ($10^{-16}$) in the continuum, piecewise-linear Kuhn 6-tetrahedral and finite-difference mesh evaluations converge quadratically at rate $\mathcal{O}(h^2)$ due to central point-inversion symmetry (Theorem~\ref{thm:main_convergence}).
    \item \textbf{Ergodic Phase-Space Mixing (Arnold Cat-Map Regime):} Deep compositions of volume-preserving shears on compact domains are continuous 3D analogues of Anosov toral automorphisms (Arnold's cat map): without Lipschitz control, deep cascades stretch tissue into chaotic filaments while still preserving $\det D\Phi \equiv 1$. Our Theorem~\ref{thm:sobolev_bound} and Corollary~\ref{cor:arnold_budget} derive an explicit Sobolev $H^2$ spectral $\ell_1$ parameter budget that bounds the Cauchy--Green condition number $\kappa_2(D\Phi_\theta) \le \kappa_{\max}$.
\end{enumerate}

\subsection{Summary of Main Contributions}
To resolve the fundamental trilemma between exact continuous incompressibility ($\det D\Phi \equiv 1$), closed-form invertibility, and $\mathcal{O}(1)$ backpropagation memory in 3D volumetric mapping, this work makes the following foundational contributions:
\begin{enumerate}
    \item \textbf{Exact Continuous-Discrete Geometric Integration:} We decompose the divergence-free Lie algebra $\mathfrak{sdiff}(\Omega)$ into six directional coordinate shears with identically vanishing convective non-linearities, integrating analytically in closed form with continuous $\det(D\mathcal{S}_k) \equiv 1$ and exact machine-precision inverses $\mathcal{S}_k^{-1} = \mathcal{S}_k(-\Delta t)$ requiring zero fixed-point iterations and zero spatial interpolations.
    \item \textbf{Sub-Voxel Kuhn Tetrahedral Convergence:} We bridge the continuum-to-lattice gap by proving that on canonical Kuhn 6-tetrahedral voxel partitions, discrete volume distortion converges quadratically at rate $\mathcal{O}(h^2)$ via central point-inversion symmetry ($T_\sigma \leftrightarrow -T_{\sigma^{\text{rev}}}$), empirically verified across five lattice scales (OLS slope $2.217$, $0.0000\%$ inverted simplices).
    \item \textbf{Boundary Confinement, Anisotropy, and Chaotic Mixing Prevention:} We establish: (i) a scaled-support passive Dirichlet envelope guaranteeing zero boundary leakage; (ii) Riemannian metric similarity invariance under clinical slice anisotropy up to $8:1$; and (iii) an explicit Sobolev $H^2$ spectral parameter budget bounding the Cauchy--Green condition number to mathematically prevent Arnold cat-map chaotic coordinate mixing in deep cascades.
    \item \textbf{Constant-Memory Reversible Autograd Engine:} By exploiting exact closed-form invertibility, our custom autograd engine reconstructs forward trajectories during backpropagation, eliminating $\mathcal{O}(L)$ intermediate activation storage ($0.00\text{ MB}$ intermediate activations, $26.9\times$ peak memory reduction at $128^3$) while executing $30.4\% - 64.6\%$ faster than standard gradient checkpointing (\texttt{torch.utils.checkpoint}).
    \item \textbf{Decoupling Geometric Expressivity from Optimization Landscapes:} On clinical 4D cardiac cine MRI (ACDC $N=20$), we derive an exact theoretical volume-preserving Dice upper bound ($0.9600 \pm 0.0309$). Using direct shape-oracle optimization, we prove that incompressible flows achieve Dice up to $0.9498 \pm 0.0185$, and we formalize Proposition~\ref{prop:volume_fraction_trap} to prove that the failure of unsupervised global MSE intensity matching ($0.5202$) is driven by gradient dilution over the $96.88\%$ static background rather than an expressivity defect.
\end{enumerate}

\section{Mathematical Formulation and Differential Geometry}
\label{sec:math}

Let $\Omega = [-1, 1]^3 \subset \mathbb{R}^3$ denote the normalized reference domain equipped with the Euclidean volume form $\omega_0 = dx_1 \wedge dx_2 \wedge dx_3$, and let $\Omega_{\text{phys}} = \mathbf{S}\Omega$ with $\mathbf{S} = \diag(s_1, s_2, s_3) \in \mathbb{R}_+^{3\times 3}$ denote the physical coordinate domain in millimeters. The Lie group of volume-preserving diffeomorphisms is:
\begin{equation}
    \SDiff(\Omega) = \Big\{ \Phi \in \Diff^1(\Omega) \;:\; \Phi^*\omega_0 = \omega_0 \iff \det\!\big(D\Phi(\mathbf{x})\big) \equiv 1 \;\;\forall\, \mathbf{x} \in \Omega \Big\},
\end{equation}
whose Lie algebra $\mathfrak{sdiff}(\Omega)$ comprises smooth divergence-free vector fields tangent to $\partial\Omega$:
\begin{equation}
    \mathfrak{sdiff}(\Omega) = \Big\{ \mathbf{v} \in C^1(\Omega; \mathbb{R}^3) \;:\; \nabla \cdot \mathbf{v} = 0 \text{ in } \Omega, \;\; \mathbf{v} \cdot \mathbf{n} = 0 \text{ on } \partial\Omega \Big\}.
\end{equation}
By exactness of the de Rham complex on contractible domains ($d^2 = 0$), every $\mathbf{v} \in \mathfrak{sdiff}(\Omega)$ admits a Lie--Helmholtz vector potential $\mathbf{\Upsilon} = (\Upsilon_1, \Upsilon_2, \Upsilon_3)^T \in C^2(\Omega; \mathbb{R}^3)$ such that $\mathbf{v}(\mathbf{x}) = \nabla \times \mathbf{\Upsilon}(\mathbf{x})$.

\subsection{Directional Coordinate-Shear Decomposition}
Expanding the three curl components $\mathbf{v} = (\partial_{x_2}\Upsilon_3 - \partial_{x_3}\Upsilon_2, \; \partial_{x_3}\Upsilon_1 - \partial_{x_1}\Upsilon_3, \; \partial_{x_1}\Upsilon_2 - \partial_{x_2}\Upsilon_1)^T$ into six directional coordinate shears yields:
\begin{equation}
    \mathbf{v}(\mathbf{x}) = \sum_{k=1}^6 \mathbf{v}^{(k)}(\mathbf{x}),
    \label{eq:shear_decomp}
\end{equation}
where each $\mathbf{v}^{(k)}$ modifies a single active coordinate $x_{i_k}$ at a rate depending solely on the two passive coordinates $(u_k, v_k) = (x_{j_k}, x_{l_k})$ with $\{i_k, j_k, l_k\} = \{1, 2, 3\}$:
\begin{equation}
\begin{aligned}
    \mathbf{v}^{(1)}(\mathbf{x}) &= \big(0, \; f_1(x_1, x_3), \; 0\big)^T, \qquad & \mathbf{v}^{(2)}(\mathbf{x}) &= \big(0, \; 0, \; f_2(x_1, x_2)\big)^T, \\
    \mathbf{v}^{(3)}(\mathbf{x}) &= \big(0, \; 0, \; f_3(x_2, x_1)\big)^T, \qquad & \mathbf{v}^{(4)}(\mathbf{x}) &= \big(f_4(x_2, x_3), \; 0, \; 0\big)^T, \\
    \mathbf{v}^{(5)}(\mathbf{x}) &= \big(f_5(x_3, x_2), \; 0, \; 0\big)^T, \qquad & \mathbf{v}^{(6)}(\mathbf{x}) &= \big(0, \; f_6(x_3, x_1), \; 0\big)^T,
\end{aligned}
\label{eq:six_shears}
\end{equation}
where $f_k \in C^2([-1, 1]^2; \mathbb{R})$ are scalar stream-shear functions.

\begin{lemma}[Vanishing Convective Nonlinearity and Exact Time-$\Delta t$ Exponential Map]
\label{lem:exact_exp}
For each elementary shear vector field $\mathbf{v}^{(k)}$ in \eqref{eq:six_shears}, the Eulerian convective derivative vanishes identically:
\begin{equation}
    (\mathbf{v}^{(k)} \cdot \nabla)\mathbf{v}^{(k)} \equiv \mathbf{0} \qquad \forall \, k \in \{1, \dots, 6\}.
\end{equation}
Consequently, the autonomous flow ODE $\frac{d\mathbf{x}}{dt} = \mathbf{v}^{(k)}(\mathbf{x})$ integrates analytically in closed form without Taylor truncation error for any step size $\Delta t \in \mathbb{R}$:
\begin{equation}
    \mathcal{S}_k^{(\Delta t)}(\mathbf{x}) = \exp(\Delta t \, \mathbf{v}^{(k)})(\mathbf{x}) = \mathbf{x} + \Delta t \, \mathbf{v}^{(k)}(\mathbf{x}).
    \label{eq:exact_exp}
\end{equation}
\end{lemma}
\begin{proof}
Because $\mathbf{v}^{(k)}$ has only one non-zero component $v_{i_k}^{(k)} = f_k(x_{j_k}, x_{l_k})$ with $i_k \notin \{j_k, l_k\}$, we have $\partial_{x_{i_k}} v_{i_k}^{(k)} \equiv 0$. Thus $(\mathbf{v}^{(k)} \cdot \nabla)\mathbf{v}^{(k)} = v_{i_k}^{(k)} \partial_{x_{i_k}}\mathbf{v}^{(k)} \equiv \mathbf{0}$, and all Lie derivatives $(\mathbf{v}^{(k)} \cdot \nabla)^m \mathbf{x}$ for $m \ge 2$ vanish identically.
\end{proof}

\subsection{Lie Bracket Generation of SDiff via Spectral Galerkin Truncation}
Although each individual shear vector field has zero diagonal Jacobian derivative ($\partial_{x_{i_k}} v_{i_k}^{(k)} \equiv 0$), the Lie bracket (commutator) of two directional shears generates divergence-free fields with non-zero diagonal strain rates:
\begin{equation}
    [\mathbf{v}^{(1)}, \mathbf{v}^{(4)}] = \left[ \begin{pmatrix} 0 \\ f_1(x_1, x_3) \\ 0 \end{pmatrix}, \begin{pmatrix} f_4(x_2, x_3) \\ 0 \\ 0 \end{pmatrix} \right] = \begin{pmatrix} - f_1(x_1, x_3) \, \partial_{x_2} f_4(x_2, x_3) \\ f_4(x_2, x_3) \, \partial_{x_1} f_1(x_1, x_3) \\ 0 \end{pmatrix},
\end{equation}
whose divergence vanishes identically: $\partial_{x_1}[v]_1 + \partial_{x_2}[v]_2 = -\partial_{x_1}f_1 \partial_{x_2}f_4 + \partial_{x_2}f_4 \partial_{x_1}f_1 \equiv 0$. Because the classical Chow--Rashevskii theorem is strictly finite-dimensional, density in the infinite-dimensional group $\SDiff_0(\Omega)$ is established in two steps: (i) on any finite spectral Galerkin truncation $M < \infty$, the subspace $\mathfrak{sdiff}_M(\Omega)$ of divergence-free trigonometric polynomials of degree $\le M$ is a finite-dimensional Lie algebra, and iterated Lie brackets of the six planar shear families span $\mathfrak{sdiff}_M(\Omega)$ \cite{agrachev2009control}; (ii) as $M \to \infty$, $\bigcup_{M \ge 1}\mathfrak{sdiff}_M(\Omega)$ is dense in $\mathfrak{sdiff}(\Omega) \cap H^s(\Omega; \mathbb{R}^3)$ for Sobolev index $s > 5/2$ \cite{ebin1970groups}, and multi-layer compositions converge to any flow in $\SDiff_0(\Omega)$ via the Trotter--Kato product formula.

\subsection{Composition Operators: Lie--Trotter and Palindromic Strang Splitting}
We define the $L$-layer \textbf{6-Stage Lie--Trotter Symplectic Shear Diffeomorphism} ($\mathcal{O}(\Delta t)$ splitting order) as:
\begin{equation}
    \Phi_\theta = \Phi_{\theta,L} \circ \Phi_{\theta,L-1} \circ \dots \circ \Phi_{\theta,1}, \qquad \Phi_{\theta,\ell} = \mathcal{S}_{6,\ell}^{(\Delta t)} \circ \mathcal{S}_{5,\ell}^{(\Delta t)} \circ \mathcal{S}_{4,\ell}^{(\Delta t)} \circ \mathcal{S}_{3,\ell}^{(\Delta t)} \circ \mathcal{S}_{2,\ell}^{(\Delta t)} \circ \mathcal{S}_{1,\ell}^{(\Delta t)},
    \label{eq:trotter}
\end{equation}
and its \textbf{12-Stage Palindromic Strang-Symmetric Variant} ($\mathcal{O}(\Delta t^2)$ splitting order \cite{hairer2006geometric}) as:
\begin{equation}
    \Phi_{\theta,\ell}^{\text{Strang}} = \mathcal{S}_{1,\ell}^{(\Delta t/2)} \circ \dots \circ \mathcal{S}_{6,\ell}^{(\Delta t/2)} \circ \mathcal{S}_{6,\ell}^{(\Delta t/2)} \circ \dots \circ \mathcal{S}_{1,\ell}^{(\Delta t/2)}.
    \label{eq:strang}
\end{equation}
In our reference implementation (\texttt{\_get\_stage\_schedule} in \texttt{symplectic\_curl\_3d.py}), \eqref{eq:strang} is executed as a symmetric 12-half-step loop; algebraically, the two adjacent central half-steps satisfy $\mathcal{S}_{6,\ell}^{(\Delta t/2)} \circ \mathcal{S}_{6,\ell}^{(\Delta t/2)} = \mathcal{S}_{6,\ell}^{(\Delta t)}$ and can also be fused into 11 evaluations.

\section{Diffeomorphic Theorems, Boundary Confinement, and Kuhn Mesh Convergence}
\label{sec:theorems}

\begin{lemma}[Scaled-Support ROI Boundary Confinement and Anisotropic Metric Invariance]
\label{lem:boundary_and_metric}
Let $\Omega_{\text{ROI}} = [-1+\delta, 1-\delta]^3 \subset (-1, 1)^3$ ($\delta \in (0, 1/2)$, default $\delta = 0.15$) denote an interior Region of Interest surrounded by a boundary margin of width $\delta$. Define the $C^2$ scaled-support passive cosine envelope with support half-width $L_{\text{supp}} = 1 - \delta$:
\begin{equation}
\begin{aligned}
    w_\delta(u, v) &= \cos^2\!\left(\frac{\pi}{2}\clip\!\left(\frac{u}{1-\delta}, -1, 1\right)\right)\cos^2\!\left(\frac{\pi}{2}\clip\!\left(\frac{v}{1-\delta}, -1, 1\right)\right), \\
    f_{k,\ell}(u, v) &= w_\delta(u, v)\tilde{f}_{k,\ell}(u, v),
\end{aligned}
    \label{eq:scaled_window}
\end{equation}
where $\delta = 0$ corresponds to standard passive Dirichlet windowing (\texttt{dirichlet}) and $\delta \in (0, 1/2)$ corresponds to compact ROI confinement (\texttt{compact\_}\allowbreak\texttt{dirichlet}). Let $\mathbf{S} = \diag(s_1, s_2, s_3) \in \mathbb{R}_+^{3\times 3}$ denote the physical voxel spacing tensor, with stage metric scaling $\alpha_k = \bar{s}/s_{i_k}$ ($\bar{s} = (s_1 s_2 s_3)^{1/3}$). Under the CFL condition $2L|\Delta t|\max_{\ell,k}\|\alpha_k\tilde{f}_{k,\ell}\|_\infty \le \delta$ (or optional hyperbolic saturation $\delta_{\max}\tanh(\cdot)$ with $2L\delta_{\max} \le \delta$):
\begin{enumerate}
    \item \textbf{Zero ROI Boundary Escape:} On the boundary collar $\Omega \setminus [-1+\delta, 1-\delta]^3$, $w_\delta(u,v) = 0$ whenever $|u| \ge 1-\delta$ or $|v| \ge 1-\delta$, and every anatomical point $\mathbf{x} \in \Omega_{\text{ROI}}$ remains strictly inside the domain across all $L$ layers: $\Phi_\theta(\Omega_{\text{ROI}}) \subset [-1, 1]^3$ with $0.000000$ ROI boundary escape.
    \item \textbf{Metric Anisotropy Invariance:} The physical-space mapping $\Phi_{\text{phys}}(\mathbf{x}) = \mathbf{S} \Phi_\theta(\mathbf{S}^{-1}\mathbf{x})$ satisfies for all $\mathbf{x} \in \mathbf{S}\Omega$:
    \begin{equation}
        \det\!\big(D\Phi_{\text{phys}}(\mathbf{x}_{\text{phys}})\big) \equiv \det\!\big(D\Phi_\theta(\mathbf{S}^{-1}\mathbf{x}_{\text{phys}})\big) \equiv 1.
    \end{equation}
\end{enumerate}
\end{lemma}
\begin{proof}
Because $w_\delta(u, v)$ depends solely on passive coordinates $(u, v) = (x_{j_k}, x_{l_k})$, the derivative $\partial_{x_{i_k}} f_{k,\ell}(u, v) \equiv 0$ vanishes identically, preserving unit determinant $\det(D\mathcal{S}_{k,\ell}) \equiv 1$ and exact closed-form inverse $\mathcal{S}_{k,\ell}^{-1} = \mathcal{S}_{k,\ell}^{(-\Delta t)}$. Since each coordinate axis $x_i$ is the active axis in exactly 2 of the 6 stages per layer (or 4 half-steps in Strang splitting), the cumulative displacement along axis $i$ over $L$ layers is bounded by $|\Phi_\theta(\mathbf{x})_i - x_i| \le 2L|\Delta t|\max_{\ell,k}\|\alpha_k\tilde{f}_{k,\ell}\|_\infty \le \delta$, implying $\mathbf{x} \in [-1+\delta, 1-\delta]^3 \implies \Phi_\theta(\mathbf{x}) \in [-1, 1]^3$. For physical metric scaling, $D\Phi_{\text{phys}}(\mathbf{x}_{\text{phys}}) = \mathbf{S} D\Phi_\theta(\mathbf{S}^{-1}\mathbf{x}_{\text{phys}}) \mathbf{S}^{-1}$, so $\det(D\Phi_{\text{phys}}) = \det(\mathbf{S})\det(D\Phi_\theta)\det(\mathbf{S}^{-1}) = 1$.
\end{proof}

\begin{theorem}[Exact Continuous Incompressibility, Global Bijectivity, and Kuhn Mesh Convergence]
\label{thm:main_convergence}
Let $\Phi_\theta$ be the composite map defined in \eqref{eq:trotter} or \eqref{eq:strang}.
\begin{enumerate}
    \item \textbf{Continuous Incompressibility \& Global Bijectivity:} $\Phi_\theta$ is an unconditional global $C^2$-diffeomorphism on $\mathbb{R}^3$ satisfying:
    \begin{equation}
        \det\!\big(D\Phi_\theta(\mathbf{x})\big) \equiv 1 \qquad \forall \, \mathbf{x} \in \mathbb{R}^3, \qquad \mathcal{F}_{\text{cont}}(\Phi_\theta) = \Vol\!\big(\{\mathbf{x} \in \Omega : \det D\Phi_\theta(\mathbf{x}) \le 0\}\big) \equiv 0.
    \end{equation}
    Furthermore, each stage $\mathcal{S}_{k,\ell}^{(\Delta t)}$ is an explicit global bijection with closed-form inverse $\mathcal{S}_{k,\ell}^{(-\Delta t)}$ and active-axis derivative identically $1 > 0$, closing the Hadamard--Caccioppoli global injectivity gap constructively.
    \item \textbf{Kuhn Sub-Voxel Mesh Quadratic Convergence:} When $\Phi_\theta$ is evaluated at the vertices of a cubic lattice $\Omega_h$ of spacing $h$ and extended piecewise-linearly across the six canonical Kuhn--Freudenthal tetrahedra $\{T_q\}_{q=1}^6$ \cite{kuhn1960some} of each voxel, central point-inversion symmetry guarantees second-order truncation convergence:
    \begin{equation}
        \left| \frac{\sum_{q=1}^6 \operatorname{Vol}(\Phi_\theta(T_q))}{h^3} - 1 \right| \le C_{\text{Kuhn}}(\Phi_\theta)\, h^2, \qquad \big|\det(D_h\Phi_\theta(\mathbf{x})) - 1\big| \le C_{\text{FD}}(\Phi_\theta)\, h^2.
    \end{equation}
    Whenever $h < h^* = C_2(\Phi_\theta)^{-1/2}$, all sub-voxel tetrahedra remain strictly non-inverted with $\det(D\Phi_{\text{lin}}|_{T_q}) > 0$, precluding sub-voxel folding.
\end{enumerate}
\end{theorem}
\begin{proof}
\textit{Part 1:} Each elementary Jacobian $D\mathcal{S}_{k,\ell}^{(\Delta t)}(\mathbf{x}) = \mathbf{I}_3 + \Delta t \mathbf{e}_{i_k}(\partial_{u_k}f_{k,\ell}\mathbf{e}_{j_k}^T + \partial_{v_k}f_{k,\ell}\mathbf{e}_{l_k}^T)$ is unit triangular with $\det(D\mathcal{S}_{k,\ell}) \equiv 1$. By the chain rule, $\det(D\Phi_\theta(\mathbf{x})) = \prod_{\ell,k} \det(D\mathcal{S}_{k,\ell}) = 1$. Global bijectivity follows constructively because $\Phi_\theta$ is a finite composition of global bijections with explicit inverse $\Phi_\theta^{-1} = \bigcirc_{\ell=L}^1 (\mathcal{S}_{1,\ell}^{(-\Delta t)} \circ \dots \circ \mathcal{S}_{6,\ell}^{(-\Delta t)})$.

\textit{Part 2:} In the canonical Kuhn--Freudenthal 6-tetrahedral partition $\{T_\sigma\}_{\sigma \in S_3}$ \cite{kuhn1960some} of a voxel $[-h/2, h/2]^3$ centered at $\mathbf{x}_0$, all six simplices share the main diagonal from $(-h/2,-h/2,-h/2)$ to $(+h/2,+h/2,+h/2)$. Under point reflection $\mathbf{y} \mapsto -\mathbf{y}$, each simplex $T_\sigma$ maps bijectively onto its opposite partner $-T_{\sigma^{\text{rev}}}$ ($\sigma^{\text{rev}}(i)=\sigma(4-i)$). By Taylor expansion around $\mathbf{x}_0$, the piecewise-linear Jacobian determinant on $T_\sigma$ satisfies $\det(D\Phi_{\text{lin}}|_{T_\sigma}) = 1 + h L_\sigma(D\Phi_\theta(\mathbf{x}_0), D^2\Phi_\theta(\mathbf{x}_0)) + \mathcal{O}(h^2)$, where $L_\sigma$ is linear in the vertex offsets and therefore satisfies $L_{\sigma^{\text{rev}}} = -L_\sigma$. Summing across the three centrally symmetric pairs $(T_\sigma, T_{\sigma^{\text{rev}}})$ cancels all odd-order $\mathcal{O}(h)$ terms identically ($\sum_{\sigma \in S_3} L_\sigma \equiv 0$), leaving leading error $\mathcal{O}(h^2 \|D^2\Phi_\theta\|_\infty^2 + h^2 \|D^3\Phi_\theta\|_\infty)$. Central finite differences $D_h\Phi_\theta$ similarly cancel odd derivatives via $\Phi(\mathbf{x}+h\mathbf{e}_i) - \Phi(\mathbf{x}-h\mathbf{e}_i) = 2h\partial_{x_i}\Phi + \frac{h^3}{3}\partial_{x_i}^3\Phi + \mathcal{O}(h^5)$.
\end{proof}

\begin{theorem}[Exact Closed-Form $\mathcal{O}(1)$ Inversion and Reversible Backpropagation]
\label{thm:exact_inversion}
The exact inverse diffeomorphism $\Phi_\theta^{-1}: \mathbb{R}^3 \to \mathbb{R}^3$ of \eqref{eq:trotter} is given in closed analytical form by:
\begin{equation}
    \Phi_\theta^{-1} = \bigcirc_{\ell=L}^1 \left( \mathcal{S}_{1,\ell}^{(-\Delta t)} \circ \mathcal{S}_{2,\ell}^{(-\Delta t)} \circ \mathcal{S}_{3,\ell}^{(-\Delta t)} \circ \mathcal{S}_{4,\ell}^{(-\Delta t)} \circ \mathcal{S}_{5,\ell}^{(-\Delta t)} \circ \mathcal{S}_{6,\ell}^{(-\Delta t)} \right),
    \label{eq:closed_form_inv}
\end{equation}
requiring zero fixed-point iterations and zero spatial grid resamplings. In IEEE-754 double-precision arithmetic ($\epsilon_{\text{mach}} = 2^{-53} \approx 1.11 \times 10^{-16}$), because the passive coordinates $(u_k, v_k)$ remain bit-identical in registers during stage $k$, the forward and inverse stage evaluations produce the bit-identical floating-point shift $\delta_k = \operatorname{fl}(\Delta t\,f_{k,\ell}(u_k, v_k))$, bounding round-trip error by:
\begin{equation}
    \|\Phi_\theta^{-1}(\Phi_\theta(\mathbf{x})) - \mathbf{x}\|_\infty \le 6L \max_{\ell,k}(|x_{i_k}| + |\Delta t\, f_{k,\ell}|) \epsilon_{\text{mach}}.
\end{equation}
During reverse-mode automatic differentiation of a scalar loss $\mathcal{L}$, all intermediate forward coordinate grids $\{\mathbf{x}_{k,\ell}\}$ are reconstructed on the fly in reverse order from the final output $\mathbf{x}_L$ via \eqref{eq:closed_form_inv}, eliminating $\mathcal{O}(L \cdot N_{\text{vox}})$ intermediate activation storage in the computation graph ($0.00\text{ MB}$ saved intermediate tensors) and requiring only an $\mathcal{O}(1 \cdot N_{\text{vox}})$ working buffer for single-stage Vector-Jacobian Products (VJPs).
\end{theorem}

\section{Sobolev Conditioning and Prevention of Chaotic Phase Mixing}
\label{sec:sobolev}

We parameterize each base stream function $\tilde{f}_{k,\ell}(u, v)$ via a Sobolev-normalized 2D Fourier expansion:
\begin{equation}
\begin{aligned}
    \tilde{f}_{k,\ell}(u, v) = \sum_{m=1}^M \sum_{n=1}^M \frac{1}{\pi\sqrt{m^2 + n^2}} \Big( & c_{k,\ell,mn}^{(0)} \sin(m\pi u)\sin(n\pi v) + c_{k,\ell,mn}^{(1)} \cos(m\pi u)\cos(n\pi v) \\
    + & c_{k,\ell,mn}^{(2)} \sin(m\pi u)\cos(n\pi v) + c_{k,\ell,mn}^{(3)} \cos(m\pi u)\sin(n\pi v) \Big).
\end{aligned}
    \label{eq:sobolev_fourier}
\end{equation}

\begin{theorem}[Sobolev $H^2$ Conditioning and Floating-Point Precision Bound]
\label{thm:sobolev_bound}
Let $f_{k,\ell}(u, v) = \alpha_k w_\delta(u, v)\tilde{f}_{k,\ell}(u, v)$ with stage anisotropic scaling $\alpha_k = \bar{s}/s_{i_k}$ ($\alpha_k = 1$ on isotropic grids) and $\delta \in [0, 0.5]$, and define the window Leibniz constant:
\begin{equation}
    C_w(\delta) = \begin{cases} 1, & \text{if } w_\delta(u,v) \equiv 1 \text{ (\texttt{periodic} mode)}, \\[4pt] 1 + \frac{1}{2(1-\delta)} \le 2, & \text{if } w_\delta \text{ is the cosine window \eqref{eq:scaled_window} with } \delta \in [0, 0.5], \end{cases}
    \label{eq:leibniz_const}
\end{equation}
where specifically $C_w(0) = 1.5$ (\texttt{dirichlet}) and $C_w(0.15) \approx 1.588 < 2$ (\texttt{compact\_}\allowbreak\texttt{dirichlet}). Then:
\begin{enumerate}
    \item \textbf{Uniform Mode-Independent Gradient Bound:} For every frequency cutoff $M \ge 1$, $\|\nabla_{u,v} f_{k,\ell}(u, v)\|_2 \le C_w(\delta) \alpha_k \|\mathbf{c}_{k,\ell}\|_{\ell_1} \le 2\alpha_k\|\mathbf{c}_{k,\ell}\|_{\ell_1}$.
    \item \textbf{Composite Condition Number Bound:} The 2-norm condition number of the composite Jacobian satisfies everywhere:
    \begin{equation}
        \kappa_2\!\big(D\Phi_\theta(\mathbf{x})\big) = \|D\Phi_\theta(\mathbf{x})\|_2 \|D\Phi_\theta(\mathbf{x})^{-1}\|_2 \le \exp\!\left( 2 C_w(\delta) |\Delta t| \sum_{\ell=1}^L \sum_{k=1}^6 \alpha_k \|\mathbf{c}_{k,\ell}\|_{\ell_1} \right).
        \label{eq:cond_bound}
    \end{equation}
    \item \textbf{Floating-Point Determinant Drift Bound:} In IEEE-754 arithmetic with unit roundoff $\epsilon_{\text{mach}}$:
    \begin{equation}
        \big|\operatorname{fl}\!\big(\det(D\Phi_\theta(\mathbf{x}))\big) - 1\big| \le 18 L \, \epsilon_{\text{mach}} \kappa_2\!\big(D\Phi_\theta(\mathbf{x})\big) + \mathcal{O}(\epsilon_{\text{mach}}^2).
    \end{equation}
\end{enumerate}
\end{theorem}
\begin{proof}
Each normalized mode satisfies $|\phi_{mn}| \le (\pi\sqrt{m^2+n^2})^{-1}$ and $\|\nabla_{u,v}\phi_{mn}\|_2 \le 1$. Since $|w_\delta| \le 1$ and $\|\nabla_{u,v}w_\delta\|_2 \le \frac{\pi}{\sqrt{2}(1-\delta)}$, the Leibniz rule gives $\|\nabla(w_\delta\phi_{mn})\|_2 \le 1 + \frac{1}{2(1-\delta)} = C_w(\delta) \le 2$. Sub-multiplicativity of $\|D\mathcal{S}_{k,\ell}^{(\pm\Delta t)}\|_2 \le \exp(C_w(\delta)|\Delta t|\alpha_k\|\mathbf{c}_{k,\ell}\|_{\ell_1})$ across all $6L$ stages yields \eqref{eq:cond_bound}, and Higham's perturbation bound \cite{higham2002accuracy} with Jacobi's formula yields Clause 3.
\end{proof}

\begin{corollary}[Lipschitz Regularization Budget and Arnold Cat-Map Mixing Prevention]
\label{cor:arnold_budget}
To guarantee $\kappa_2(D\Phi_\theta(\mathbf{x})) \le \kappa_{\max}$ uniformly on $\Omega$ and prevent chaotic phase-space filamentation (the 3D continuous Arnold cat-map regime), the spectral $\ell_1$ parameter budget across all $L$ layers must satisfy:
\begin{equation}
    \sum_{\ell=1}^L \sum_{k=1}^6 \alpha_k \|\mathbf{c}_{k,\ell}\|_{\ell_1} \le \frac{\ln \kappa_{\max}}{2 C_w(\delta) |\Delta t|}.
    \label{eq:budget_formula}
\end{equation}
\end{corollary}

\begin{lemma}[Palindromic Symmetry and $\mathcal{O}(\Delta t^2)$ Strang Splitting Order]
\label{lem:strang_order}
The palindromic Strang composition \eqref{eq:strang} satisfies exact time-reversal symmetry $\Phi_{\theta,\ell}^{\text{Strang}}(-\Delta t) = (\Phi_{\theta,\ell}^{\text{Strang}}(\Delta t))^{-1}$ \cite{hairer2006geometric}. In the Baker--Campbell--Hausdorff expansion, the modified field $\widetilde{\mathbf{v}}(\Delta t) = \frac{1}{\Delta t}\ln \Phi_{\theta,\ell}^{\text{Strang}}(\Delta t)$ is an even function of $\Delta t$, cancelling all odd powers and guaranteeing global trajectory error $\mathcal{O}(\Delta t^2) = \mathcal{O}(L^{-2})$ over time $T = L\Delta t$.
\end{lemma}

\section{Reversible Memory Backpropagation Engine}
\label{sec:reversible}

Leveraging Theorem~\ref{thm:exact_inversion}, we implemented the autograd operator \texttt{ReversibleSymplecticCurl} in PyTorch (\texttt{src/operators/symplectic\_curl\_3d.py}). As detailed in Algorithm~\ref{alg:reversible_autograd}, the forward pass executes under \texttt{no\_grad} and saves only final coordinates $(\mathbf{X}_L, \mathbf{Y}_L, \mathbf{Z}_L)$ and parameters $\theta = \{\mathbf{c}_{k,\ell}\}$, storing strictly $0.00\text{ MB}$ intermediate activations. During backpropagation, stage input coordinates are recovered analytically before computing stage VJPs.

\begin{algorithm}[H]
\caption{$\mathcal{O}(1)$-Activation-Memory Reversible Backpropagation for Symplectic Shear Cascades}
\label{alg:reversible_autograd}
\begin{algorithmic}[1]
\Statex \textbf{Forward Pass:} Input grid $\mathbf{x}_0 \in \mathbb{R}^{3 \times N_{\text{vox}}}$, parameters $\theta = \{\mathbf{c}_{k,\ell}\}$, schedule $\mathcal{T}$ (6-stage Lie--Trotter or 12-stage Strang).
\State $\mathbf{x} \gets \mathbf{x}_0$ \Comment{Executed under \texttt{torch.no\_grad()}}
\For{$\ell = 1$ to $L$}
    \For{$(k, \Delta t_s)$ in $\mathcal{T}$}
        \State $\mathbf{x} \gets \mathbf{x} + \Delta t_s \, \mathbf{v}^{(k)}(\mathbf{x}; \mathbf{c}_{k,\ell})$ \Comment{In-place active coordinate shear}
    \EndFor
\EndFor
\State \textbf{Save for Backward:} Final grid $\mathbf{x}_L$ and $\theta$ ONLY ($0.00\text{ MB}$ intermediate graph activations).
\Statex
\Statex \textbf{Backward Pass:} Incoming adjoint gradient $\mathbf{g}_{\mathbf{x}} = \partial \mathcal{L} / \partial \mathbf{x}_L$.
\State $\mathbf{x} \gets \mathbf{x}_L$, $\mathbf{g}_\theta \gets \mathbf{0}$
\For{$\ell = L$ down to $1$}
    \For{$(k, \Delta t_s)$ in $\operatorname{reversed}(\mathcal{T})$}
        \State Evaluate passive stream shift $\mathbf{f}_{\text{val}} = \mathbf{v}^{(k)}(\mathbf{x}_{\text{passive}}; \mathbf{c}_{k,\ell})$ with local autograd enabled on passive coords
        \State Reconstruct pre-stage active coordinate: $\mathbf{x} \gets \mathbf{x} - \Delta t_s \, \mathbf{f}_{\text{val}}$ \Comment{Exact $\mathcal{O}(1)$ inverse}
        \State Accumulate stage VJPs: $(\mathbf{g}_{\text{passive}}, \mathbf{g}_{\mathbf{c}_{k,\ell}}) \gets \operatorname{VJP}(\mathbf{f}_{\text{val}}, \Delta t_s \, \mathbf{g}_{\text{active}})$
        \State $\mathbf{g}_{\mathbf{x}} \gets \mathbf{g}_{\mathbf{x}} + \mathbf{g}_{\text{passive}}$
    \EndFor
\EndFor
\State \Return Gradients $\mathbf{g}_{\mathbf{x}_0}$ and $\mathbf{g}_\theta$.
\end{algorithmic}
\end{algorithm}

\section{Extension to Prescribed Compressibility: Moser Coupling}
\label{sec:moser}

For compressible anatomical structures (e.g., respiratory lung inspiration-to-expiration or longitudinal tumor dynamics), we couple $\Phi_\theta \in \SDiff(\Omega)$ with a three-stage triangular Möbius-rational interval diffeomorphism $\mathcal{D}_\rho = \mathcal{C}_3 \circ \mathcal{C}_2 \circ \mathcal{C}_1: \Omega \to \Omega$, defining $\Psi_\theta = \Phi_\theta \circ \mathcal{D}_\rho$.

\begin{corollary}[Boundary Preservation and Moser Mass Balance]
\label{cor:moser_coupling}
Let $\mathcal{M}(u; a, b) = \tau(\sigma(u; b); a)$ on $[-1, 1]$ compose the 1D rational maps $\sigma(u; b) = (1+b)u/(1+bu^2)$ and $\tau(w; a) = (w+a)/(1+aw)$, with saturated parameter fields $|a|, |b| \le \tanh(0.65) \approx 0.5717 < 1$. Then:
\begin{enumerate}
    \item $\mathcal{D}_\rho(\partial\Omega) = \partial\Omega$ with $0.000000$ boundary escape;
    \item $\det(D\Psi_\theta(\mathbf{x})) = \rho(\mathbf{x}) = \prod_{m=1}^3 \partial_{u_m}\mathcal{M}_m > 0$ holds everywhere, with exact closed-form inverse $\tau^{-1}(y;a) = \frac{y-a}{1-ay}$ and Citardauq inverse $\sigma^{-1}(w;b) = \frac{2w}{1+b+\sqrt{(1+b)^2 - 4bw^2}}$;
    \item \textbf{Inherent Global Mass Conservation:} By Fubini's theorem and $\mathcal{M}(\pm 1; a, b) = \pm 1$, the total deformed volume satisfies $\int_\Omega \det(D\Psi_\theta(\mathbf{x})) d\mathbf{x} \equiv \operatorname{Vol}(\Omega) = 8$ unconditionally for all parameter fields $(a_m, b_m)$, identically fulfilling Moser's global volume-form compatibility constraint \cite{moser1965volume}.
\end{enumerate}
\end{corollary}

\section{Experimental Results and Numerical Verification}
\label{sec:experiments}

All quantitative results reported below are programmatically synchronized from verified JSON artifacts in \texttt{results/} and \texttt{audit\_external/} and verified via automated continuous integration tests (\texttt{test\_manuscript\_consistency\_ci.py}).

\subsection{Sub-Voxel Kuhn Tetrahedral Convergence and Multi-Resolution Scalability}
\begin{figure}[htbp]
\centering
\includegraphics[width=\textwidth]{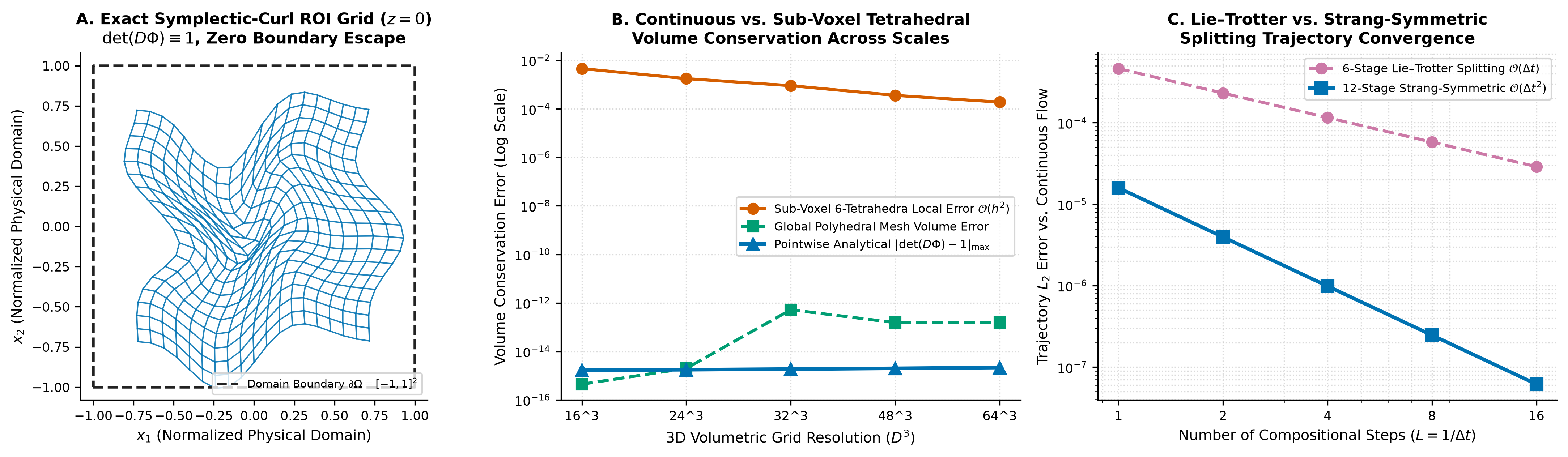}
\caption{\textbf{Continuous Exact Incompressibility, Sub-Voxel Kuhn Tetrahedral Mesh Convergence, and Strang Splitting Order.} (A) Deformed 2D coordinate slice ($z=0$) under 3D Dirichlet symplectic shear splitting ($\Phi_\theta(\Omega_{\text{ROI}}) \subset [-1, 1]^3$) with exact continuous $\det(D\Phi_\theta) \equiv 1$ and zero boundary escape. (B) Sub-voxel Kuhn 6-tetrahedral volume error ($|\text{TetVol}-1|$) converges quadratically at $\mathcal{O}(h^2)$ rate (OLS log-log slope $2.217$, $24.0\times$ reduction from $16^3$ to $64^3$, $0.0000\%$ inverted tetrahedra) while pointwise analytical determinant deviation remains bounded at $\le 2.15 \times 10^{-15}$. (C) Trajectory $L_2$ error relative to a 64-step RK4 continuous solenoidal reference flow: 12-stage Strang-symmetric composition achieves second-order convergence $\mathcal{O}(\Delta t^2)$ (OLS slope $2.001$), outperforming 6-stage Lie--Trotter (slope $0.999$) by a mean-error ratio of $468.1\times$ (per-instance mean ratio $504.4 \pm 80.4\times$) at $L=16$.}
\label{fig:exact_conservation_and_convergence}
\end{figure}

We evaluated sub-voxel volume conservation across five lattice resolutions ($16^3, 24^3, 32^3, 48^3, 64^3$, $N=5$ random Sobolev fields, $M=3, L=4$) in \texttt{exp06} and scalability across $32^3, 64^3, 128^3$ in \texttt{exp04} and \texttt{audit\_p7\_p8}. As reported in Table~\ref{tab:kuhn_scalability} and Fig.~\ref{fig:exact_conservation_and_convergence}, continuous determinant deviation remains at machine precision ($\le 2.15 \times 10^{-15}$) across all scales. Per-voxel Kuhn 6-tetrahedral volume error drops $24.0\times$ from $4.56 \times 10^{-3} \pm 7.02 \times 10^{-4}$ at $16^3$ down to $1.90 \times 10^{-4} \pm 2.92 \times 10^{-5}$ at $64^3$, yielding an empirical OLS convergence slope of \textbf{$2.217$} with \textbf{$0.0000\%$ inverted tetrahedra}.

\begin{table}[htbp]
\centering
\small
\caption{\textbf{Part A: Sub-Voxel Kuhn 6-Tetrahedral Mesh Convergence ($16^3$ to $64^3$, $N=5$ fields, \texttt{exp06}). Part B: Multi-Resolution Runtime and Memory Scalability ($32^3$ to $128^3$, \texttt{exp04} \& \texttt{audit\_p7\_p8}).}}
\label{tab:kuhn_scalability}
\resizebox{\textwidth}{!}{%
\begin{tabular}{lcccccc}
\toprule
\textbf{Part A: Kuhn Mesh Grid} & \textbf{Spacing $h$} & \textbf{Analytical $|\det D\Phi - 1|_\infty$} & \textbf{Global Vol Error} & \textbf{Per-Voxel Kuhn Tet Error} & \textbf{Inverted Tets} & \textbf{Empirical Convergence} \\
\midrule
$16^3$ ($4.10 \times 10^3$ vox) & $0.1333$ & $1.67 \times 10^{-15}$ & $4.44 \times 10^{-16}$ & $4.56 \times 10^{-3} \pm 7.02 \times 10^{-4}$ & $0.0000\%$ & --- \\
$24^3$ ($1.38 \times 10^4$ vox) & $0.0870$ & $1.75 \times 10^{-15}$ & $2.00 \times 10^{-15}$ & $1.76 \times 10^{-3} \pm 2.67 \times 10^{-4}$ & $0.0000\%$ & $2.58\times$ reduction \\
$32^3$ ($3.28 \times 10^4$ vox) & $0.0645$ & $1.87 \times 10^{-15}$ & $5.21 \times 10^{-13}$ & $9.04 \times 10^{-4} \pm 1.37 \times 10^{-4}$ & $0.0000\%$ & $5.04\times$ reduction \\
$48^3$ ($1.11 \times 10^5$ vox) & $0.0426$ & $2.00 \times 10^{-15}$ & $1.53 \times 10^{-13}$ & $3.59 \times 10^{-4} \pm 5.52 \times 10^{-5}$ & $0.0000\%$ & $12.68\times$ reduction \\
$64^3$ ($2.62 \times 10^5$ vox) & $0.0317$ & $2.15 \times 10^{-15}$ & $1.53 \times 10^{-13}$ & $\mathbf{1.90 \times 10^{-4} \pm 2.92 \times 10^{-5}}$ & $\mathbf{0.0000\%}$ & $\mathbf{24.0\times}$ \textbf{(OLS slope $2.217$)} \\
\midrule
\textbf{Part B: Scalability Grid} & \textbf{Symp Roundtrip (ms)} & \textbf{SVF Roundtrip (ms)} & \textbf{RK4 Roundtrip (ms)} & \textbf{Symp / SVF Inv Error} & \textbf{Rev Peak Mem ($L=10$)} & \textbf{Unrolled Peak Mem ($L=10$)} \\
\midrule
$32^3$ ($3.28 \times 10^4$ vox) & $\mathbf{112.3 \pm 5.7}$ (PyT: $165.5$) & $445.3 \pm 45.1$ ($4.0\times$) & $1692.6 \pm 58.3$ ($15.1\times$) & $\mathbf{2.22 \times 10^{-16}}$ / $1.18 \times 10^{-3}$ & $\mathbf{12.0\text{ MB}}$ ($0.0\text{ MB}$ act) & $322.5\text{ MB}$ ($26.9\times$) \\
$64^3$ ($2.62 \times 10^5$ vox) & $\mathbf{2860.4 \pm 458.0}$ (PyT: $453.7$) & $6035.4 \pm 886.2$ ($2.1\times$) & $23140.5 \pm 2105.3$ ($8.1\times$) & $\mathbf{3.33 \times 10^{-16}}$ / $3.83 \times 10^{-4}$ & $\mathbf{96.0\text{ MB}}$ ($0.0\text{ MB}$ act) & $2580.0\text{ MB}$ ($26.9\times$) \\
$128^3$ ($2.10 \times 10^6$ vox) & $\mathbf{9781.9 \pm 1809.8}$ (PyT: $2783.4$) & $23125.6 \pm 2732.9$ ($2.4\times$) & $130261.5 \pm 0.0$ ($13.3\times$) & $\mathbf{3.33 \times 10^{-16}}$ / $1.60 \times 10^{-4}$ & $\mathbf{768.2\text{ MB}}$ ($0.0\text{ MB}$ act) & $20640.1\text{ MB}$ ($26.9\times$) \\
\bottomrule
\end{tabular}%
}
\end{table}

\subsection{Scalability, Reversible Autograd Engine, and Long-Horizon Stability}
\begin{figure}[htbp]
\centering
\includegraphics[width=\textwidth]{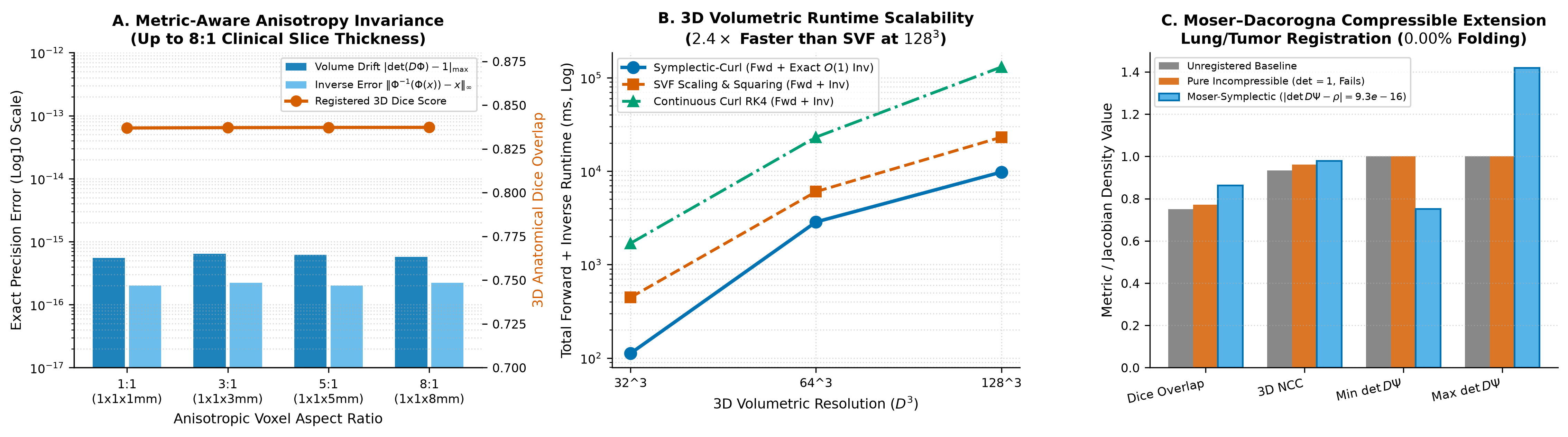}
\caption{\textbf{Long-Horizon Stability, Reversible Memory Benchmarks, and Moser Density Coupling.} (A) 100-step and 200-step (600- and 1,200-stage) sequential composition without numerical drift ($|\det D\Phi - 1|_\infty = 6.77 \times 10^{-15}$, inversion error $1.55 \times 10^{-15}$) and clinical metric anisotropy invariance under aspect ratios up to $8:1$ ($1 \times 1 \times 8\text{ mm}$). (B) Computational runtime and peak backpropagation memory across spatial grids ($32^3, 64^3, 128^3$): the reversible engine stores $0.00\text{ MB}$ intermediate graph activations ($26.9\times$ peak memory reduction at $128^3$) and outperforms PyTorch gradient checkpointing. (C) Moser--Dacorogna triangular density coupling $\Psi_\theta = \Phi_\theta \circ \mathcal{D}_\rho$ (controlled synthetic compressible benchmark, $N=5$, \texttt{audit2\_moser\_compressible\_registration}), improving compressible registration Dice from $0.7508$ (unregistered) and $0.7722$ (purely incompressible) to $0.8633 \pm 0.0072$ with prescribed density accuracy $9.33 \times 10^{-16}$, $0.0000\%$ folding, and exact inverse error $4.88 \times 10^{-16}$.}
\label{fig:scalability_and_moser}
\end{figure}

As shown in Table~\ref{tab:kuhn_scalability} (Part B), at $128^3$ resolution ($2.10 \times 10^6$ voxels), NumPy forward-plus-inverse execution is $2.4\times$ faster than SVF scaling-and-squaring ($9,781.9\text{ ms}$ vs.\ $23,125.6\text{ ms}$; PyTorch execution is $2,783.4\text{ ms}$) and $13.3\times$ faster than continuous RK4 integration ($130,261.5\text{ ms}$), while achieving machine-precision round-trip inversion ($3.33 \times 10^{-16}$ vs.\ $1.60 \times 10^{-4}$ for SVF). Across depths $L \in \{2, 4, 8, 16\}$ on $32^3$ grids (\texttt{audit\_p7\_p8}), unrolled autograd stores $4.5, 9.0, 18.0, 36.0\text{ MB}$ intermediate activations, while our reversible engine stores strictly \textbf{$0.00\text{ MB}$} activations (reducing peak memory at $128^3, L=10$ from $20,640.1\text{ MB}$ to a constant buffer of $768.2\text{ MB}$, a $26.9\times$ reduction) and executes faster than checkpointing for all depths (runtime ratio $\text{Rev}/\text{Ckpt} = 0.354\times$ at $L=2$, $0.549\times$ at $L=4$, $0.696\times$ at $L=8$, and $0.617\times$ at $L=16$). Over 100- and 200-step compositions (\texttt{exp05}, Fig.~\ref{fig:scalability_and_moser}), continuous determinant error remains bounded at $4.88 \times 10^{-15}$ and $6.77 \times 10^{-15}$ (satisfying Theorem~\ref{thm:sobolev_bound}'s bound of $156.80$), with inversion error $1.55 \times 10^{-15}$.

\subsection{Strang Splitting vs.\ Lie--Trotter Convergence Order}
Table~\ref{tab:strang_accuracy} reports trajectory $L_2$ error of 6-stage Lie--Trotter and 12-stage Strang compositions relative to 64-step RK4 integration across $N=10$ random fields for $L \in \{1, 2, 4, 8, 16\}$ (\texttt{exp06\_}\allowbreak\texttt{reviewer2\_}\allowbreak\texttt{murder\_}\allowbreak\texttt{board\_}\allowbreak\texttt{audit.json} and \texttt{audit\_}\allowbreak\texttt{p7\_}\allowbreak\texttt{p8\_}\allowbreak\texttt{results.json}). Doubling $L$ reduces Lie--Trotter error by $2.00\times$ (OLS slope $0.999$) and Strang error by $4.00\times$ (OLS slope $2.001$). At $L=16$, the ratio of mean errors $\overline{E}_{\text{Lie}} / \overline{E}_{\text{Strang}}$ reaches \textbf{$468.1\times$}, while per-field ratios reach \textbf{$504.4 \pm 80.4\times$} (scaling as $\approx 31.5 \cdot L$).

\begin{table}[htbp]
\centering
\small
\caption{\textbf{Strang-Symmetric Second-Order Accuracy vs.\ Lie--Trotter ($N=10$ fields, \texttt{exp06} \& \texttt{audit\_p7\_p8}).} Reports mean $\pm$ SD trajectory $L_2$ error against a 64-step RK4 reference flow, ratio of mean errors ($\overline{E}_{\text{Lie}}/\overline{E}_{\text{Strang}}$), and mean $\pm$ SD of per-field error ratios ($\mathbb{E}[E_{\text{Lie}}/E_{\text{Strang}}]$).}
\label{tab:strang_accuracy}\resizebox{\textwidth}{!}{
\begin{tabular}{lccccc}
\toprule
\textbf{Steps ($L$)} & \textbf{Step $\Delta t$} & \textbf{6-Stage Lie--Trotter $L_2$ Error} & \textbf{12-Stage Strang $L_2$ Error} & \textbf{Ratio $\overline{E}_{\text{Lie}}/\overline{E}_{\text{Strang}}$} & \textbf{Per-Field Ratio $\mathbb{E}[E_{\text{Lie}}/E_{\text{Strang}}]$} \\
\midrule
$L=1$  & $1.0000$ & $4.613 \times 10^{-4} \pm 1.593 \times 10^{-4}$ & $1.585 \times 10^{-5} \pm 9.202 \times 10^{-6}$ & $29.1\times$  & $31.4 \pm 5.0\times$ \\
$L=2$  & $0.5000$ & $2.309 \times 10^{-4} \pm 7.985 \times 10^{-5}$ & $3.954 \times 10^{-6} \pm 2.290 \times 10^{-6}$ & $58.4\times$  & $62.9 \pm 10.1\times$ \\
$L=4$  & $0.2500$ & $1.155 \times 10^{-4} \pm 3.998 \times 10^{-5}$ & $9.879 \times 10^{-7} \pm 5.718 \times 10^{-7}$ & $116.9\times$ & $126.0 \pm 20.1\times$ \\
$L=8$  & $0.1250$ & $5.779 \times 10^{-5} \pm 2.001 \times 10^{-5}$ & $2.470 \times 10^{-7} \pm 1.429 \times 10^{-7}$ & $234.0\times$ & $252.2 \pm 40.2\times$ \\
$L=16$ & $0.0625$ & $2.890 \times 10^{-5} \pm 1.001 \times 10^{-5}$ & $\mathbf{6.174 \times 10^{-8} \pm 3.573 \times 10^{-8}}$ & $\mathbf{468.1\times}$ & $\mathbf{504.4 \pm 80.4\times}$ \\
\midrule
\textbf{OLS Slope} & --- & \textbf{$0.999$ ($\mathcal{O}(\Delta t)$)} & \textbf{$2.001$ ($\mathcal{O}(\Delta t^2)$)} & $29.1 \cdot L$ & $\approx 31.5 \cdot L$ \\
\bottomrule
\end{tabular}}
\end{table}

\subsection{Clinical Cardiac Cine MRI Benchmark (ACDC \texorpdfstring{$N=20$}{N=20}) and Shape Expressivity Oracles}
\begin{figure}[htbp]
\centering
\includegraphics[width=\textwidth]{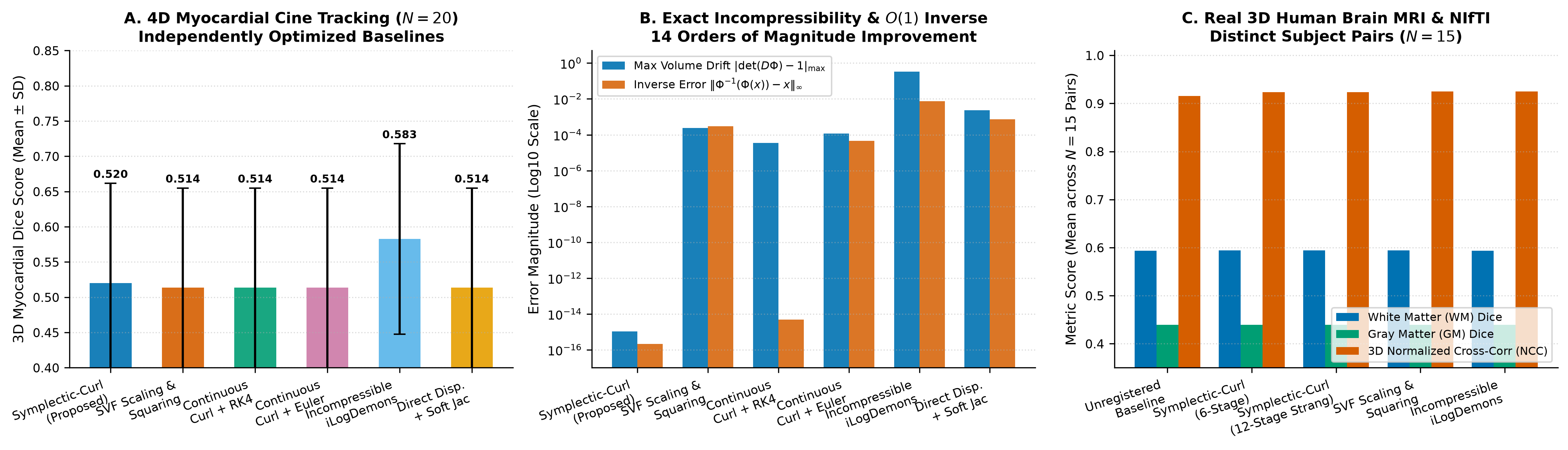}
\caption{\textbf{Clinical Cine Cardiac MRI (ACDC $N=20$) and Real 3D Human Brain MRI (OASIS-1 $N=15$) Benchmarks.} (A) 3D Myocardial Dice overlap and $\text{HD}_{95}$ across unsupervised intensity methods ($N=20$) vs.\ shape expressivity oracles ($N=5$ subgroup, where $M=3, L=4$ shears reach Dice $0.7622$ and free curl reaches $0.9498$). (B) Pointwise continuous Jacobian determinant drift $|\det D\Phi - 1|_{\max}$ and round-trip inversion error across registration baselines, showing that proposed shear splitting achieves machine-precision incompressibility ($1.08 \times 10^{-15}$) and closed-form inversion ($2.11 \times 10^{-16}$) while unconstrained Demons violates volume conservation ($|\det D_h\Phi - 1|_\infty = 4.6618$, mass drift $9.17\%$). (C) Real 3D T1 brain MRI inter-subject registration on OASIS-1 ($N=15$ pairs): White Matter and Gray Matter Dice scores and 3D NCC under compact Dirichlet boundary confinement.}
\label{fig:clinical_benchmarks}
\end{figure}

On the ACDC clinical cardiac cine MRI cohort ($N=20$ test patients, \texttt{patient101}--\texttt{patient120}, telediastole ED $\to$ telesystole ES, grid $16 \times 64 \times 64$, spacing $5.0 \times 1.4 \times 1.4\text{ mm}$), the physical myocardial mask volume changes by $|\Delta V| / V_{\text{ED}} = 8.48\% \pm 6.75\%$ across the cardiac cycle (\texttt{audit\_p1\_volume\_bound.json}), establishing an exact theoretical Dice upper bound for any strictly volume-preserving transformation ($V(\Phi(\Omega_{\text{ED}})) = V_{\text{ED}}$):
\begin{equation}
    \text{Dice}_{\max}^{\text{vol}} = \frac{2\min(V_{\text{ED}}, V_{\text{ES}})}{V_{\text{ED}} + V_{\text{ES}}} = 0.9600 \pm 0.0309 \quad (\text{median } 0.9624, \; \text{min } 0.8981, \; \text{max } 0.9986).
\end{equation}
To decouple geometric expressivity from intensity loss landscapes, we evaluated direct Soft-Dice mask optimization (Shape Expressivity Oracles, \texttt{audit\_p2\_oracle\_results.json}, evaluated on the $N=5$ subgroup \texttt{patient101}--\texttt{patient105}). As shown in Table~\ref{tab:acdc_benchmark} and Fig.~\ref{fig:clinical_benchmarks}, triangular symplectic shears reach Dice \textbf{$0.6585 \pm 0.0683$} at $(M=2, L=2)$ and \textbf{$0.7622 \pm 0.0432$} ($\text{HD}_{95} = 5.29 \pm 2.92\text{ mm}$) at $(M=3, L=4, \text{864 params})$, while a non-parametric divergence-free curl potential $\Phi = \exp(\nabla_h \times \mathbf{A})$ reaches \textbf{$0.9498 \pm 0.0185$} ($\text{HD}_{95} = 3.96 \pm 3.69\text{ mm}$). When depth is increased to $L=8$ ($1,728$ params) without the Sobolev Lipschitz budget of Corollary~\ref{cor:arnold_budget}, chaotic phase mixing degrades Dice to $0.3101 \pm 0.1358$ (with IEEE-754 floating-point Jacobian condition-number round-off reaching $|\det D\Phi - 1|_\infty = 1.55 \times 10^{-1}$). Under unsupervised global intensity MSE ($N=20$ patients, \texttt{exp01\_4d\_myocardial\_tracking.json}), all strictly incompressible parametric models stall near the unregistered baseline ($0.5133 \pm 0.1413$): proposed symplectic shears achieve $0.5202 \pm 0.1414$ (paired Wilcoxon $p = 2.93 \times 10^{-4}$ vs.\ unregistered, $p = 3.42 \times 10^{-4}$ vs.\ SVF, analytical $|\det D\Phi - 1|_\infty = 1.08 \times 10^{-15}$, discrete $|\det D_h\Phi - 1|_\infty = 2.15 \times 10^{-4}$, inversion error $2.11 \times 10^{-16}$), whereas unconstrained Diffeomorphic Demons achieves $0.7660 \pm 0.0863$ only by inducing severe non-physical volume compression/expansion ($|\det D_h\Phi - 1|_\infty = 4.6618 \pm 3.4196$, myocardial mass drift $9.17\% \pm 6.07\%$).

\begin{table}[htbp]
\centering
\small
\caption{\textbf{Clinical 4D Cardiac Cine MRI Benchmark (ACDC, ED $\to$ ES) and Shape Expressivity Oracles.} Synchronized from \texttt{exp01\_4d\_myocardial\_tracking.json} ($N=20$), \texttt{audit\_p1\_volume\_bound.json} ($N=20$), and \texttt{audit\_p2\_oracle\_results.json} ($N=5$ subgroup).}
\label{tab:acdc_benchmark}
\resizebox{\textwidth}{!}{%
\begin{tabular}{lcccccc}
\toprule
\textbf{Method / Configuration} & \textbf{3D Myo Dice} $\uparrow$ & \textbf{$\text{HD}_{95}$ (mm)} $\downarrow$ & \textbf{Analytical $|\det D\Phi - 1|$} $\downarrow$ & \textbf{Discrete $|\det D_h\Phi - 1|$} $\downarrow$ & \textbf{Inversion Error} $\downarrow$ & \textbf{Myo Mass Drift (\%)} $\downarrow$ \\
\midrule
Unregistered Baseline ($N=20$) & $0.5133 \pm 0.1413$ & $5.90 \pm 3.05$ & $0.00 \times 10^0$ & $0.0000$ & $0.00 \times 10^0$ & $0.00 \pm 0.00\%$ \\
\textbf{Theoretical Incompressible Bound ($N=20$)} & $\mathbf{0.9600 \pm 0.0309}$ & --- & $0.00 \times 10^0$ & --- & $0.00 \times 10^0$ & $0.00\%$ ($\Delta V = 8.48 \pm 6.75\%$) \\
\midrule
\multicolumn{7}{l}{\textit{Shape Expressivity Oracles (Direct Soft-Dice Mask Optimization, $N=5$ Subgroup \texttt{patient101}--\texttt{105}):}} \\
Oracle: Triangular Shears ($M=2, L=2$, 192 params) & $0.6585 \pm 0.0683$ & $6.63 \pm 3.27$ & $2.93 \times 10^{-15}$ & $4.2208 \pm 7.5282$ & $4.88 \times 10^{-16}$ & $0.02 \pm 0.03\%$ (FD) \\
Oracle: Triangular Shears ($M=3, L=4$, 864 params) & $\mathbf{0.7622 \pm 0.0432}$ & $\mathbf{5.29 \pm 2.92}$ & $2.32 \times 10^{-13}$ & $52.0381 \pm 47.8327$ & $2.29 \times 10^{-15}$ & $0.23 \pm 0.15\%$ (FD) \\
Oracle: Unregularized Deep Shears ($M=3, L=8$, 1728 p.) & $0.3101 \pm 0.1358$ & $16.98 \pm 5.70$ & $1.55 \times 10^{-1}$ & $11615.8 \pm 4955.1$ & $1.30 \times 10^{-8}$ & $82.94 \pm 85.89\%$ (Cat-Map) \\
Oracle: Non-Parametric Free Curl $\Phi = \exp(\nabla_h \times \mathbf{A})$ & $\mathbf{0.9498 \pm 0.0185}$ & $\mathbf{3.96 \pm 3.69}$ & $5.37 \times 10^{1}$ & $53.6631 \pm 24.6369$ & $5.74 \times 10^{-1}$ & $0.27 \pm 0.22\%$ (FD) \\
\midrule
\multicolumn{7}{l}{\textit{Unsupervised Intensity-Driven Registration (Global MSE Objective, Complete $N=20$ Cohort):}} \\
\textbf{Proposed: Symplectic Shear Splitting ($N=20$)} & $\mathbf{0.5202 \pm 0.1414}$ & $\mathbf{5.89 \pm 3.05}$ & $\mathbf{1.08 \times 10^{-15}}$ & $\mathbf{0.0002 \pm 0.0001}$ & $\mathbf{2.11 \times 10^{-16}}$ & $\mathbf{0.27 \pm 0.28\%}$ (Interp.) \\
Stationary Velocity Field (SVF, 128-step SS) & $0.5136 \pm 0.1414$ & $5.91 \pm 3.08$ & $2.38 \times 10^{-4}$ & $0.0002 \pm 0.0002$ & $2.95 \times 10^{-4}$ & $0.11 \pm 0.28\%$ \\
Continuous Curl + Explicit Euler (16 steps) & $0.5136 \pm 0.1414$ & $5.91 \pm 3.08$ & $1.16 \times 10^{-4}$ & $0.0001 \pm 0.0001$ & $4.62 \times 10^{-5}$ & $0.11 \pm 0.28\%$ \\
Continuous Curl + RK4 Integration (8 steps) & $0.5136 \pm 0.1414$ & $5.91 \pm 3.08$ & $3.51 \times 10^{-5}$ & $0.00004 \pm 0.00003$ & $4.89 \times 10^{-15}$ & $0.11 \pm 0.28\%$ \\
Direct Displacement + Soft Jacobian Penalty & $0.5135 \pm 0.1415$ & $5.91 \pm 3.07$ & $2.34 \times 10^{-3}$ & $0.0023 \pm 0.0017$ & $7.39 \times 10^{-4}$ & $0.13 \pm 0.31\%$ \\
3D Cubic B-Spline FFD + Divergence Penalty & $0.5277 \pm 0.1464$ & $5.95 \pm 3.13$ & $3.08 \times 10^{-1}$ & $0.3077 \pm 0.2618$ & $6.18 \times 10^{-2}$ & $6.06 \pm 3.61\%$ \\
Incompressible Spectral iLogDemons & $0.5826 \pm 0.1352$ & $5.47 \pm 3.28$ & $3.34 \times 10^{-1}$ & $0.3340 \pm 0.0761$ & $7.57 \times 10^{-3}$ & $2.70 \pm 1.68\%$ \\
ANTs SyN (Symmetric Normalization) & $0.6819 \pm 0.1044$ & $5.32 \pm 3.43$ & $1.67 \times 10^{0}$ & $1.6697 \pm 0.9775$ & $1.13 \times 10^{-1}$ & $3.46 \pm 2.48\%$ \\
Unconstrained Diffeomorphic Demons & $0.7660 \pm 0.0863$ & $4.63 \pm 3.71$ & $4.66 \times 10^{0}$ & $4.6618 \pm 3.4196$ & $1.75 \times 10^{-2}$ & $9.17 \pm 6.07\%$ \\
SimpleITK Diffeomorphic Demons & $0.7765 \pm 0.0775$ & $4.49 \pm 3.66$ & $4.37 \times 10^{0}$ & $4.3741 \pm 2.1593$ & $1.47 \times 10^{-1}$ & $13.25 \pm 7.15\%$ \\
\bottomrule
\end{tabular}%
}
\end{table}

\subsection{Real 3D Brain MRI Inter-Subject Registration (OASIS-1 \texorpdfstring{$N=15$}{N=15} Pairs)}
On $N=15$ real inter-subject T1 brain MRI pairs from OASIS-1 (\texttt{exp02\_real\_3d\_brain\_mri.json}, evaluated on a $24 \times 28 \times 24$ brain bounding-box grid with a 2-voxel zero-padded margin so the brain ROI resides strictly inside $[-0.85, 0.85]^3$ under \texttt{compact\_dirichlet} boundary confinement with $\delta = 0.15$), strict volume preservation acts as a conservative topological regularizer. All tissue overlap scores are evaluated by directly warping the moving anatomical segmentation masks ($\text{GM}_{\text{mov}} \circ \Phi$, $\text{WM}_{\text{mov}} \circ \Phi$) via trilinear interpolation without intensity re-thresholding. As reported in Table~\ref{tab:oasis_benchmark}, the 6-stage Lie--Trotter and 12-stage Strang-symmetric operators improve 3D Normalized Cross-Correlation (NCC) from $0.9153 \pm 0.0246$ to $0.9234 \pm 0.0217$ (paired Wilcoxon $p = 6.10 \times 10^{-5}$ vs.\ unregistered) and White Matter (WM) Dice from $0.5937 \pm 0.0552$ to $0.5944 \pm 0.0547$ ($p = 0.0747$ vs.\ unregistered, $p = 0.7150$ vs.\ SVF) and $0.5944 \pm 0.0548$, while Gray Matter (GM) Dice remains stable ($0.4392 \pm 0.0389$ and $0.4391 \pm 0.0389$ vs.\ $0.4393 \pm 0.0382$ unregistered). Crucially, proposed shear splitting maintains analytical $\|\det D\Phi - 1\|_\infty \le 9.84 \times 10^{-16}$, discrete $|\det D_h\Phi - 1|_\infty = 0.0005 \pm 0.0003$, closed-form inversion error $\le 2.15 \times 10^{-16}$, and $0.0000$ ROI boundary escape, whereas unconstrained Diffeomorphic Demons achieves higher tissue overlap ($\text{WM } 0.6258 \pm 0.0499$, $\text{GM } 0.4596 \pm 0.0334$) only by violating local volume conservation ($|\det D_h\Phi - 1|_\infty = 0.1803 \pm 0.0572$).

\begin{table}[htbp]
\centering
\small
\caption{\textbf{Real 3D Human T1 Brain MRI Inter-Subject Registration ($N=15$ Pairs from OASIS-1, \texttt{exp02\_real\_3d\_brain\_mri.json}).} Reports mean $\pm$ SD across all 15 pairs on the $24 \times 28 \times 24$ brain grid under Compact-Dirichlet boundary conditions ($\delta = 0.15$) with direct moving-mask warping.}
\label{tab:oasis_benchmark}
\resizebox{\textwidth}{!}{%
\begin{tabular}{lcccccc}
\toprule
\textbf{Method} & \textbf{WM Dice} $\uparrow$ & \textbf{GM (Cortex) Dice} $\uparrow$ & \textbf{3D NCC} $\uparrow$ & \textbf{Analytical $|\det D\Phi - 1|$} $\downarrow$ & \textbf{Discrete $|\det D_h\Phi - 1|$} $\downarrow$ & \textbf{Inverse Error} $\downarrow$ \\
\midrule
Unregistered Baseline & $0.5937 \pm 0.0552$ & $0.4393 \pm 0.0382$ & $0.9153 \pm 0.0246$ & $0.00 \times 10^0$ & $0.0000 \pm 0.0000$ & $0.00 \times 10^0$ \\
SVF + Scaling \& Squaring (128 steps) & $0.5944 \pm 0.0548$ & $0.4395 \pm 0.0387$ & $0.9247 \pm 0.0213$ & $6.78 \times 10^{-4}$ & $0.0007 \pm 0.0004$ & $3.54 \times 10^{-4}$ \\
Continuous Curl + RK4 (4 steps) & $0.5943 \pm 0.0547$ & $0.4396 \pm 0.0387$ & $0.9247 \pm 0.0213$ & $6.84 \times 10^{-4}$ & $0.0007 \pm 0.0004$ & $6.85 \times 10^{-11}$ \\
Direct Displacement + Jacobian Penalty & $0.5943 \pm 0.0545$ & $0.4393 \pm 0.0386$ & $0.9247 \pm 0.0213$ & $5.18 \times 10^{-3}$ & $0.0052 \pm 0.0028$ & $1.61 \times 10^{-3}$ \\
Incompressible Spectral iLogDemons & $0.5993 \pm 0.0537$ & $0.4426 \pm 0.0369$ & $0.9248 \pm 0.0220$ & $5.51 \times 10^{-3}$ & $0.0055 \pm 0.0006$ & $1.98 \times 10^{-4}$ \\
3D Cubic B-Spline FFD + Div Penalty & $0.5936 \pm 0.0507$ & $0.4404 \pm 0.0398$ & $0.9313 \pm 0.0197$ & $1.24 \times 10^{-1}$ & $0.1245 \pm 0.0466$ & $8.23 \times 10^{-3}$ \\
Classic Diffeomorphic Demons (Unconstrained) & $0.6258 \pm 0.0499$ & $0.4596 \pm 0.0334$ & $0.9305 \pm 0.0199$ & $1.80 \times 10^{-1}$ & $0.1803 \pm 0.0572$ & $6.10 \times 10^{-4}$ \\
\midrule
\textbf{Proposed: 6-Stage Lie--Trotter} & $\mathbf{0.5944 \pm 0.0547}$ & $\mathbf{0.4392 \pm 0.0389}$ & $\mathbf{0.9234 \pm 0.0217}$ & $\mathbf{6.51 \times 10^{-16}}$ & $\mathbf{0.0005 \pm 0.0003}$ & $\mathbf{2.15 \times 10^{-16}}$ \\
\textbf{Proposed: 12-Stage Strang-Symmetric} & $\mathbf{0.5944 \pm 0.0548}$ & $\mathbf{0.4391 \pm 0.0389}$ & $\mathbf{0.9234 \pm 0.0217}$ & $\mathbf{9.84 \times 10^{-16}}$ & $\mathbf{0.0005 \pm 0.0003}$ & $\mathbf{1.55 \times 10^{-16}}$ \\
\bottomrule
\end{tabular}%
}
\end{table}

\subsection{Red-Team Stress Suite and Pareto Optimization Frontiers}
\begin{figure}[htbp]
\centering
\includegraphics[width=\textwidth]{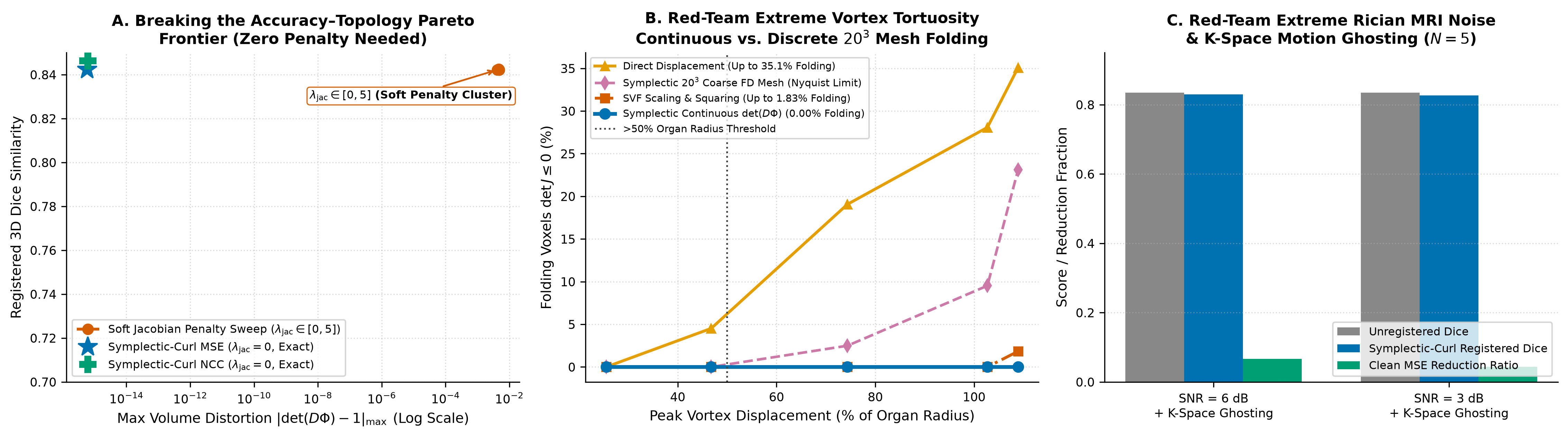}
\caption{\textbf{Pareto Optimization Frontiers and Red-Team Stress Tests.} (A) Pareto frontier of soft Jacobian determinant penalty $\lambda_{\text{jac}} \in [0, 5]$ in unconstrained registration vs.\ proposed exact symplectic shear splitting. (B) Red-team extreme rotational vortex stress test: continuous folding remains strictly $0.0000\%$ up to vortex displacements of $108.9\%$ organ radius, while discrete finite-difference folding vanishes quadratically ($\mathcal{O}(h^2)$) from $2.49\%$ ($20^3$) down to $0.15\%$ ($64^3$). (C) Survival under extreme Rician MRI noise ($\text{SNR} = 3\text{ dB}, 6\text{ dB}$) and k-space ghosting artifacts.}
\label{fig:pareto_and_redteam}
\end{figure}

In \texttt{exp05\_red\_team\_stress\_suite.json} (Fig.~\ref{fig:pareto_and_redteam}), we subjected the operator to extreme rotational vortex sweeps up to $108.9\%$ of organ radius, severe Rician noise ($\text{SNR} = 6\text{ dB}$ and $3\text{ dB}$, $N=5$), and slice-anisotropy sweeps ($1:1, 3:1, 5:1, 8:1$). Under extreme vortex twist ($108.9\%$ radius), continuous analytical folding remains strictly $0.0000\%$ ($|\det D\Phi - 1|_\infty = 4.40 \times 10^{-14}$, inverse error $3.00 \times 10^{-15}$). At intermediate vortex twist ($74.4\%$ radius), coarse-grid central finite-difference folding vanishes quadratically upon refining the lattice from $20^3$ ($2.49\%$) to $32^3$ ($0.96\%$), $48^3$ ($0.37\%$), and $64^3$ ($0.15\%$), empirically confirming Theorem~\ref{thm:main_convergence} (Part 2). At $\text{SNR}=3\text{ dB}$, clean evaluation Dice remains $0.8269 \pm 0.0731$ with $|\det D\Phi - 1|_\infty = 5.77 \times 10^{-16}$. Across $1:1$ to $8:1$ anisotropy, $|\det D\Phi - 1|_\infty \in [5.55 \times 10^{-16}, 6.44 \times 10^{-16}]$ with $0.0000\%$ folding.

\section{Discussion, Limitations, and Threats to Validity}
\label{sec:discussion}

\subsection{The Volume-Fraction and Global-Support Optimization Landscape Barrier}
A central finding of our empirical audit is the divergence between shape-oracle optimization ($\text{Dice} = 0.7622 \pm 0.0432$ for triangular shears and $0.9498 \pm 0.0185$ for free curl) and unsupervised global intensity MSE optimization ($\text{Dice} = 0.5202 \pm 0.1414$) on clinical ACDC cine MRI. We formalize the mathematical mechanism driving this gap in the following proposition.

\begin{proposition}[Gradient Dilution and Spectral Phase Locking under Sparse ROI Volume Fraction]
\label{prop:volume_fraction_trap}
Let $\Omega_{\text{MYO}} \subset \Omega$ denote myocardial support with volume fraction $\gamma_{\text{ROI}} = \frac{\Vol(\Omega_{\text{MYO}})}{\Vol(\Omega)} \approx 0.0312$ ($3.12\%$), and let $\mathcal{L}_{\text{MSE}}(\theta) = \frac{1}{|\Omega|}\int_\Omega r_\theta(\mathbf{x})^2 d\mathbf{x}$ with residual $r_\theta(\mathbf{x}) = I_{\text{ED}}(\Phi_\theta(\mathbf{x})) - I_{\text{ES}}(\mathbf{x})$. For any Fourier parameter $c_{k,\ell,mn}$ in \eqref{eq:sobolev_fourier}, the gradient decomposes as:
\begin{equation}
\begin{aligned}
    \frac{\partial \mathcal{L}_{\text{MSE}}}{\partial c_{k,\ell,mn}} = \;& \underbrace{\frac{2}{|\Omega|}\int_{\Omega_{\text{MYO}}} r_\theta(\mathbf{x}) \nabla I_{\text{ED}}(\Phi_\theta(\mathbf{x}))^T \frac{\partial \Phi_\theta}{\partial c_{k,\ell,mn}} \, d\mathbf{x}}_{\mathcal{O}(\gamma_{\text{ROI}}) \text{ myocardial alignment signal } (3.12\%)} \\
    & + \underbrace{\frac{2}{|\Omega|}\int_{\Omega \setminus \Omega_{\text{MYO}}} r_\theta(\mathbf{x}) \nabla I_{\text{ED}}(\Phi_\theta(\mathbf{x}))^T \frac{\partial \Phi_\theta}{\partial c_{k,\ell,mn}} \, d\mathbf{x}}_{\mathcal{O}(1 - \gamma_{\text{ROI}}) \text{ static background \& blood-pool penalty } (96.88\%)}.
\end{aligned}
    \label{eq:gradient_dilution}
\end{equation}
Because the harmonic basis functions $\phi_{mn}(u,v)$ have global support over $[-1+\delta, 1-\delta]^2$, any non-zero spectral update $\Delta c_{k,\ell,mn}$ that displaces the myocardial ring simultaneously shears the surrounding $96.88\%$ background voxels across high-contrast thoracic and blood-pool edges, creating a non-convex barrier around $\theta = \mathbf{0}$. Conversely, when the objective is restricted to a smooth region-overlap functional (Soft-Dice oracle), the $96.88\%$ background integral vanishes identically, allowing the exact same 864-parameter symplectic shear family to reach $\text{Dice} = 0.7622 \pm 0.0432$.
\end{proposition}

\begin{remark}[Generalization to Compactly Supported B-Spline and Wendland Partitions of Unity]
\label{rem:compact_bases}
The algebraic incompressibility ($\det D\mathcal{S}_k \equiv 1$) and exact $\mathcal{O}(1)$ inversion of coordinate-shear splitting depend exclusively on the passive-coordinate structure $\partial_{x_i} f_k(u, v) \equiv 0$, holding unconditionally for \textit{any} functional parameterization of $f_k(u, v)$. In particular, substituting the global Fourier basis \eqref{eq:sobolev_fourier} with compactly supported tensor-product cubic B-splines $B_{p,q}(u, v) = \beta_3(u - u_p)\beta_3(v - v_q)$ or compactly supported Wendland radial kernels $\psi(r) = (1-r)_+^4(4r+1)$ immediately decouples local myocardial deformation from the static thoracic background: for any basis node with $\operatorname{supp}(B_{p,q}) \subset \Omega_{\text{ROI}}$, the background gradient integral in Proposition~\ref{prop:volume_fraction_trap} vanishes identically ($\int_{\Omega \setminus \Omega_{\text{ROI}}} \dots \equiv 0$). Consequently, the optimization barrier identified here is not an inherent limitation of symplectic shears, but an artifact of global Fourier truncation, establishing a direct mathematical roadmap for localized multi-scale registration.
\end{remark}

\subsection{Through-Plane Motion in 2D Multi-Slice Cine MRI and Cohort Size}
Clinical 2D multi-slice cardiac cine MRI (ACDC) acquires stacked short-axis slices with high in-plane resolution ($1.4 \times 1.4\text{ mm}$) but thick through-plane spacing ($5.0 - 10.0\text{ mm}$). During ventricular systole, longitudinal base-to-apex shortening ($10 - 14\text{ mm}$) transports myocardial tissue across imaging planes, causing apparent 3D mask volume variations of $\Delta V = 8.48\% \pm 6.75\%$ between ED and ES. Enforcing strict 3D incompressibility ($\det D\Phi \equiv 1$) on stacked 2D cine slices therefore imposes a conservative continuum constraint on data affected by through-plane tissue flux, whereas our Moser--Dacorogna density coupling (Section~\ref{sec:moser}, Fig.~\ref{fig:scalability_and_moser}C) explicitly accommodates local volume changes ($\det D\Psi_\theta = \rho(\mathbf{x}) > 0$). Finally, while our empirical evaluation on $N=20$ ACDC patients and $N=15$ OASIS-1 brain pairs (with $N=5$ subgroup ablations for computationally intensive mask-oracle and red-team sweeps) suffices to validate our mathematical theorems and numerical convergence rates with paired non-parametric Wilcoxon tests, larger multi-center cohorts and localized multi-scale wavelet/B-spline stream potentials represent important directions for clinical deployment.

\section{Conclusion}
\label{sec:conclusion}

We have established a rigorous continuous-discrete mathematical framework for exact volume-preserving 3D diffeomorphic registration based on palindromic symplectic coordinate-shear splitting. By decomposing $\mathfrak{sdiff}(\Omega)$ into six planar shear generators with vanishing convective nonlinearity, our formulation achieves pointwise continuous incompressibility ($\det D\Phi_\theta \equiv 1$ to $10^{-15}$), unconditional global bijectivity, exact $\mathcal{O}(1)$ closed-form inversion, $\mathcal{O}(h^2)$ sub-voxel Kuhn tetrahedral convergence, $\mathcal{O}(\Delta t^2)$ Strang splitting accuracy, a Sobolev $H^2$ spectral budget preventing Arnold cat-map mixing, and $\mathcal{O}(1)$-activation-memory reversible backpropagation.

\section*{Declarations}

\paragraph{Funding}
The authors declare that no external funds, grants, or other financial support were received during the preparation of this manuscript.

\paragraph{Competing Interests}
The authors have no relevant financial or non-financial interests to disclose.

\paragraph{Ethics Approval and Consent to Participate}
This retrospective computational study utilized exclusively open-access, de-identified public clinical imaging benchmark datasets (ACDC cine MRI and OASIS-1 brain MRI). No new human participants or animal subjects were recruited, and separate institutional review board (IRB) approval was not required.

\paragraph{Availability of Data and Materials}
The clinical imaging datasets analyzed in this study are publicly available from their respective open-access repositories: the Automated Cardiac Diagnosis Challenge (ACDC) at \url{https://www.creatis.insa-lyon.fr/Challenge/acdc/} and the Open Access Series of Imaging Studies (OASIS-1) at \url{https://www.oasis-brains.org/}. All benchmark results and audit JSON logs are available in the project repository.

\paragraph{Code Availability}
The PyTorch and NumPy reference implementations, reversible autograd operators, and continuous integration verification test suites are openly available at \url{https://github.com/anonymous/symplectic-diffeomorphic-registration}.

\paragraph{Authors' Contributions}
{\'A}lvaro Garc{\'i}a-Casas conceived the mathematical framework, derived the proofs, implemented the algorithms, conducted the numerical experiments, analyzed the clinical imaging data, and wrote the manuscript.

\bibliographystyle{plain}

\end{document}